\documentclass{llncs}
\usepackage{llncsdoc}

\usepackage{multirow}
\usepackage{courier}
\usepackage{algorithm}
\usepackage{algpseudocode}
\usepackage{amsmath}
\usepackage{amsfonts}
\usepackage{subfig}
\usepackage{enumitem}
\usepackage{graphicx}
\usepackage{multicol}
\usepackage[labelfont=bf]{caption}
\usepackage{subfig}
\usepackage{float}
\usepackage{color}

\usepackage{booktabs} 
\usepackage{algorithm}
\usepackage{algpseudocode}
\usepackage{amsmath}
\usepackage{subfig}
\usepackage{amssymb}
\usepackage{graphicx}
\usepackage{enumitem}
\usepackage{textcomp }
\usepackage{wrapfig}

\usepackage{booktabs}

\usepackage{xcolor}
\usepackage{pbox}

\begin{document}
	\title{Twitter for Sparking a Movement, \\ Reddit for Sharing the Moment: \\ \#metoo through the Lens of Social Media}
    \author{Lydia Manikonda, Ghazaleh Beigi, Huan Liu, \and Subbarao Kambhampati}
\institute{\email{\{lmanikon, gbeigi, huan.liu, rao\}@asu.edu} \\Arizona State University, Tempe, Arizona, USA}
	
	\maketitle
    \vspace{-20pt}
	\begin{abstract}
Social media platforms are revolutionizing the way users communicate by increasing the exposure to highly stigmatized issues in the society. Sexual abuse is one such issue that recently took over social media via attaching the hashtag \#metoo to the shared posts. Individuals with different backgrounds and ethnicities began sharing their unfortunate personal experiences of being assaulted. Through comparative analysis of the tweets via \#meToo on Twitter versus the posts shared on the \#meToo subreddit, this paper makes an initial attempt to assess public reactions and emotions. Though nearly equal ratios of negative and positive posts are shared on both platforms, Reddit posts are focused on the sexual assaults within families and workplaces while Twitter posts are on showing empathy and encouraging others to continue the \#metoo movement. The data collected in this research and preliminary analysis demonstrate that users use various ways to share their experience, exchange ideas and encourage each other, and social media is suitable for groundswells such as \#metoo movement.
\vspace{-15pt}
	\end{abstract}

	\section{Introduction}
    \vspace{-10pt}
	Social media platforms are enabling individuals to maintain privacy and to self-disclose their true feelings and physical conditions~\cite{goffman2009stigma,altman1973social,cozby1973self,joinson2007self}. Especially, this practice is more common in scenarios such as sharing opinions about highly stigmatized topics in the society such as mental health, obesity, cancer, etc. In contrary to this, sexual abuse which has been traditionally brushed aside due to the fear of shame, retribution and retaliation, did finally see the light. Sexual abuse, and abuse in general is a very difficult topic for individuals to talk about it irrespective of whether the environment is an online or an offline setting~\cite{coffey1996mediators,finkelhor1990sexual}. In United States itself, on an average  there are 321,500 victims (age 12 or older) of rape and sexual assault each year. Ages 12-34 are the highest risk years for rape and sexual assault. The trauma of the sexual abuse has resulted in the long-term negative impacts such as anxiety, suicidal behavior, PTSD, panic disorder, psychosis, mood and behavioral disorder problems~\cite{sexualtrauma}. 
	
	Research states that disclosing the abuse, results in a positive impact  psychologically~\cite{mcclain2013female,andalibi2016understanding}. However, often times, the responses the survivors get from others through the disclosure could often lead to an additional emotional distress. Social media platforms are becoming more persuasive to safely disclose such issues. In the recent days, one movement that has rampantly exposed and is still exposing the sexual abuse of individuals is the \emph{\#metoo} movement. Although the term \emph{\#metoo} was originally coined in 2006 by social activists to raise awareness about sexual abuse, it became viral in October 2017, following the alleged sexual misconduct in the Hollywood. Shortly after, countless number of individuals from all over the world came forward sharing their personal stories or endorsing their support to this viral online movement. As a part of this movement, numerous posts with \#metoo hashtag have been shared especially on online platforms like Twitter. However it is not very clear what the individuals are sharing through these posts when they are attaching this specific hashtag. Through a comparative analysis of posts shared on Twitter and Reddit and by leveraging the machine learning approaches, this paper investigates these three main research questions: 
	\begin{enumerate}
		\item What kinds of specific sub-topics are the individuals talking about? 
		\item How are individuals labeling the sexual abuse and \#metoo movement? 
        \item Are individuals positive or being judgmental towards this viral movement? 
	\end{enumerate}
    \vspace{-20pt}
	\section{Data}
    \vspace{-10pt}
	We obtain two sets of data from Twitter and Reddit using the python APIs for Twitter\footnote{https://developer.twitter.com/} and Reddit\footnote{https://praw.readthedocs.io/en/latest/} respectively. We collect 620,348 posts from 205,489 users on Twitter and 190 posts from 70 users on Reddit. On Twitter, we crawl the posts (from October 2017 to January 2018) that are attached with the \#metoo hashtag where as for Reddit, we crawl all the self posts shared on {\em /r/metoo} subreddit. We crawl only the publicly shared posts and the data includes all the meta-data associated with the post. None of these posts are geo-tagged which suggests that individuals are focused exclusively on the content they want to share and not in sharing their geographical location.
    \vspace{-13pt}
	\section{Social Engagement}
    \vspace{-10pt}

    \begin{wraptable}{r}{5.5cm}
    \small
    \vspace{-0.65in}
		\begin{tabular}{|l|c|c|c|c|}
			\hline
			\emph{Eng. Att.} & \emph{Mean} & \emph{Min} & \emph{Max} & \emph{Std}\\ \hline
			\emph{Favorites} & 5.69 & 0 & 104464 & 229.14 \\ \hline
			\emph{Retweets} & 2.38 & 0 & 22893 & 71.79 \\ \hline
			\emph{Mentions} & 1.13 & 1 & 25 & 0.81 \\ \hline
			\emph{Hashtags} & 1.93 & 1 & 36 & 2.12 \\ \hline
		\end{tabular}
		\caption{Statistics about the engagement attributes}
		\label{tab:TwitterStats}
        \vspace{-0.35in}
	\end{wraptable}
    
	Due to the sensitivity of the topic and specifically the viral nature of the \#metoo hashtag, we first want to investigate how the tweets shared on this topic engage other users on Twitter. For this purpose, we compute statistics about the engagement attributes that include -- number of favorites these tweets received, number of times a tweet from this dataset is retweeted, number of mentions in the tweets and the number of hashtags attached to these tweets. Table~\ref{tab:TwitterStats} shows that on an average these tweets receive atleast 5 favorites and 2 retweets which is relatively more engaging compared to general tweets~\cite{manikonda2016tweeting}. This might be due to other users endorsing the tweets. On the other hand, it is surprising to see that users tend to engage in conversations with other users or atleast mention them in their tweets more prominently. 
	
\begin{wrapfigure}{r}{0.5\textwidth}
     \vspace{-12mm}
		\begin{center}
		\includegraphics[width=0.5\textwidth]{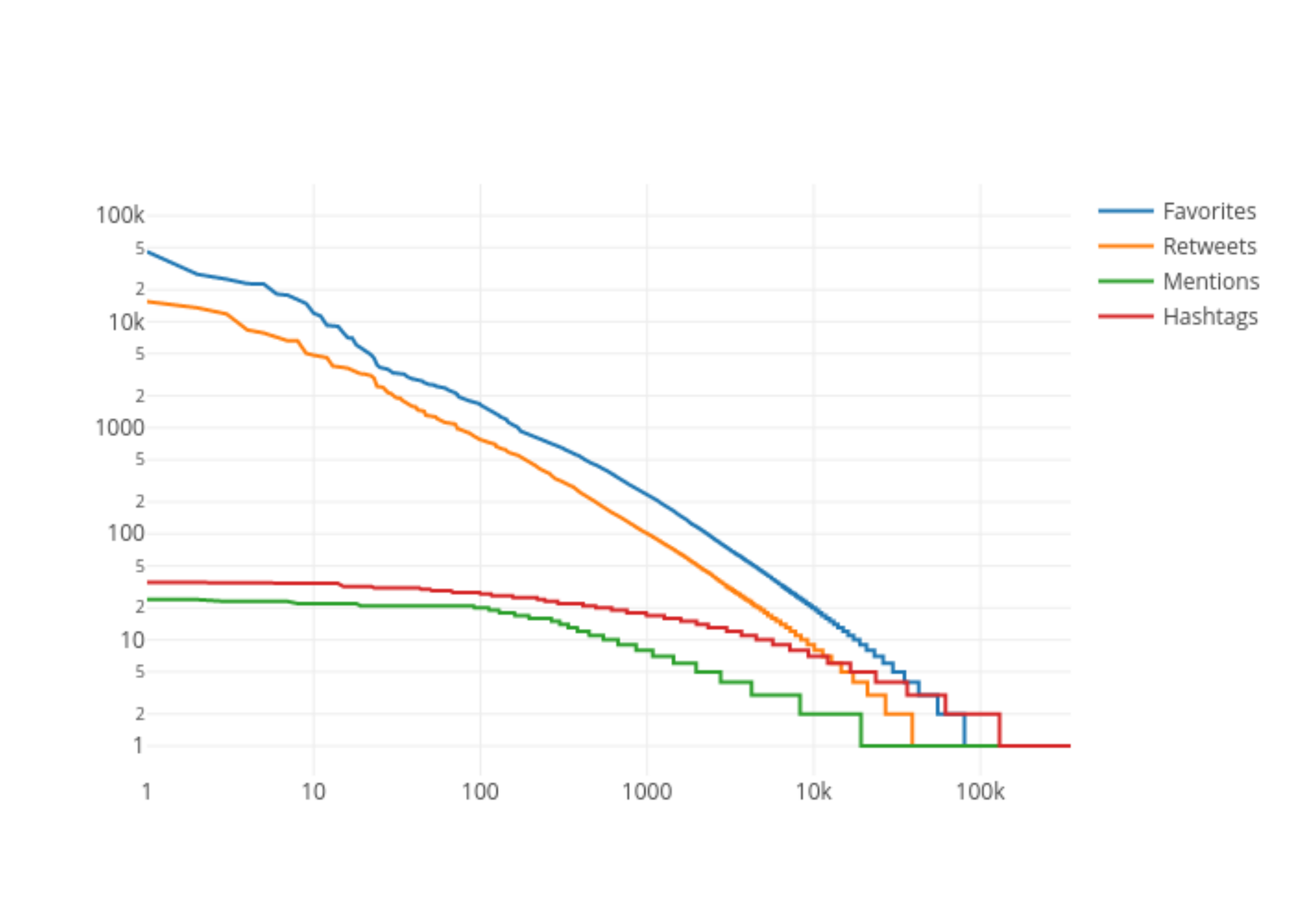}
	\end{center}
    \vspace{-8mm}
    \caption{Log-Log plot for engagement attributes on Twitter}
		\label{fig:TwitterStats}
        \vspace{-15mm}
     \end{wrapfigure}

	
Figure~\ref{fig:TwitterStats} shows the log-log plot of these engagement attributes shedding light on the favorites and retweets received by these posts. All the engagement attributes follow a power-law distribution showing that there exist few posts which are very highly engaging relative to the majority of the remaining posts. Complementing these observation, Reddit posts which can receive both up votes and down votes, receive 2.26 up votes (standard deviation=2.03) on an average. None of the {\em self} posts shared on \#metoo subreddit received down votes suggesting that posts shared on Reddit are positively engaging.
\vspace{-20pt}
	\section{Linguistic Themes to Understand the Content}
    \vspace{-10pt}
    Since \#metoo related posts are socially engaging, it is important to understand the content of these posts. We first extract the latent topics present in these posts and then focus on how users label sexual abuse through their vocabulary usage. 
	\subsection{Latent Topic Extraction}
	\vspace{-7pt}
    \begin{table}[h]
    \vspace{-30pt}
		\caption{\textbf{Topic Vocabulary.}}
        \label{tab:data_stat}
			\subfloat[\bf{Reddit}]{\begin{tabular}{l p{0.4\textwidth}}
					Topic & Top Words\\ \hline
					0&\textcolor{blue}{[Experience and memory, emotions]}  emotional, response, attacker, unwelcome, crime, severe, mugged, notmeanymore, starting, threat
					\\ \hline
					1&\textcolor{blue}{[Story Details]} date, shoulder,apartment, squad, strange, morning, van, club, escape, partner
					\\ \hline
					2&\textcolor{blue}{[Fight against harassment, being strong]} movement, victims, survivor, abused, metooers, damage, accusations, battle, strength, suffer, wounds, trigger
					\\ \hline
					3&\textcolor{blue}{[Story Details]} home, remember, hell, car, boyfriend, hotel, officer, realtor, banker, fuck, weekend, conversation
					\\ \hline
					4&\textcolor{blue}{[Offenders affiliation]} head, older, house, attorney, military, hand, working, sally, face, army, black
					\\ \hline
				\end{tabular}}
				\quad
				\subfloat[\bf{Twitter}]{\begin{tabular}{l p{0.4\textwidth}}
						
						Topic & Top Words\\ \hline
						0&\textcolor{blue}{[Fight against harassment]}  \#metoo, harassment, movement, assault, campaign, silence, violence, teaching, abuse, business, \#timesup, schools, workplace
						\\ \hline
						1&\textcolor{blue}{[Sharing news]} stories, weinstein, damon, harvey, share, backlash, mcgowan, allegations, news, misconduct, hollywood, accused, pbs
						\\ \hline
						2&\textcolor{blue}{[Sharing support by posting hashtags]}  \#metoo, \#timesup, \#goldenglobes, black, \#oprah, \#resist, winfrey, \#oprah2020, hollywood, president, \#millennials, \#veterans
						\\ \hline
						3&\textcolor{blue}{[Real story sharing]}  lewinsky, bill, clinton, monica, \#trump,accuser, video, \#feminism, \#rape, black,\#maga,simmons, russell
						\\ \hline
						4&\textcolor{blue}{[Discussing news]} witch, hunt, social, campaign, harassment, world,woody,allen,reckoning
						\\ \hline
					\end{tabular}}
	\label{LDA}
    \vspace{-20pt}
	\end{table}
	
We extract the latent topics from the corpus containing all the {\em self} posts shared on \#metoo subreddit as well as the corpus of all posts attached with the \#metoo hashtag on Twitter. Topic analysis can help to understand what aspects of sexual abuse are individuals focusing on the two platforms. We use LDA (Latent Dirichlet Allocation) topic modeling technique~\cite{blei2003latent} to extract the latent topics shown in Table~\ref{LDA}.a and~\ref{LDA}.b for Reddit and Twitter respectively.

On both these platforms, people share their unfortunate experiences of getting assaulted (topic 0 in Reddit and topic 2 in Twitter). Users also encourage each other to be strong and fight against harassment by contributing to the movement (topic 2 in Reddit and topic 0 in Twitter). However, we notice two significant difference in types of posts shared on Reddit and Twitter. On Reddit, survivors mainly share the details of the story (e.g. how and when that happened to them) and how they were hurt emotionally (topics 0, 1 and 3). They also mention the affiliation of offenders (topic 4). While On Twitter, people do not expose the details and mainly focus on supporting victims of sexual violence by just posting relevant hashtags (topic 2 summarizes the mostly used hashtag during the movement), sharing relevant news (topic 1 and 3) and urls of related news articles (topic 4). On Twitter, users also tell their stories of being harassed at workplace and how they fear being retaliated for complaining about the harassment (topic 0). These differences  between Reddit and Twitter might be because of the character limit enforced by these platforms. In particular, Reddit has allowed users to share more details and thus users might be able to reveal their real feelings easier than Twitter. But it is interesting to see that this movement became viral due to the posts shared on Twitter~\footnote{http://time.com/5051822/time-person-year-alyssa-milano-tarana-burke/}. 
    \vspace{-10pt}
    
    \subsection{Labeling Sexual Abuse}
    \subsubsection{$n$-gram Analysis}
	
    \begin{wraptable}{r}{8.5cm}
    \vspace{-0.34in}
		\small 
		\begin{tabular}{c|p{5cm}}
        \hline
        \textbf{Twitter Bigrams} & metoo movement; sexual harassment; metoo timesup; metoo campaign; metoo moment; say metoo; metoo story; social media; witch hunt; sexual misconduct \\ \hline
            \textbf{Reddit Bigrams} & sexual harassment; dont want; sexual assault; years old; dont think; Im sorry; one day; metoo movement; first time; will never \\ \hline 
            \textbf{Twitter Unigrams} & metoo; women; movement; sexual; men; harassment; now; assault; time; hollywood \\ \hline 
            \textbf{Reddit Unigrams} & men; like; me; im; women; dont; people; know; time; sexual \\ \hline 
		\end{tabular}
		\caption{Top-10 \emph{uni}-grams and \emph{bi}-grams}
		\label{tab:ngrams}
        \vspace{-0.42in}
    \end{wraptable}

To obtain a basic understanding of the content shared, we extract $n$-grams. Table~\ref{tab:ngrams} shows the $bi$-grams and $uni$-grams extracted from both the Twitter as well as Reddit posts. Bigrams show that majority of the Reddit posts focus on individual experiences about sexual harassment for example: \emph{years old}, \emph{Im sorry}, \emph{one day}, etc., where as Twitter posts focus on the existing sexual assault stories and opinions about how to address these issues (\emph{metoo movement}, \emph{say metoo}, \emph{social media}, etc). Unigrams also highlight similar set of observations. Using the most frequently occurring keywords in these corpuses, we dig little deeper to understand how users label sexual abuse through the words associated with these keywords. 
    
	\subsubsection{Considering Syntactic as well as Semantic Relationships}
    
    \vspace{-18pt} To ensure that both the syntactic and semantic relationships are captured, we represent the vocabulary of the corpus in a vector space and then measure their similarities by utilizing the popular Word2Vec approach~\cite{word2vec}. Through the pairwise word relationships shown in Table~\ref{tab:wordvec} for the most frequently occurring keywords  suggest that some of these societal entities such as {\em men} are aggressive and violate certain aspects where as {\em woman} is associated with {\em humiliated publicly} and are {\em intimidated}. Whenever users are mentioning the personal experiences ({\em story}), it is highly correlated with words such as {\em heartbreaking, frightening, terrifying, horrifying, awful,} etc. Alongside, most of the other keywords (such as {\em sex}, {\em rape}, {\em victims},  etc ) are similarly associated with a vocabulary that is mostly negative. However, the users are also recognizing that these issues should be addressed immediately (see keyword {\em timesup}) and is slightly on an encouraging side compared to other keyword relationships. Words such as {\em movement} is co-occurring with words such as {\em hysterical}, {\em nonsense} and it is not very clear if users are mocking the \#metoo movement. This may require further analysis and is out of scope for this paper. Due to the limited set of posts from Reddit, we didn't find the co-occurring patterns significant.
    \begin{table}[ht]
        \vspace{-10pt}
        \centering
    \begin{tabular}{|c|p{0.85\textwidth}|} 
    \hline
    \textbf{Keyword} & Most co-occurring words\\ \hline 
    {\em men} & aggressive, pigs, socialized, violate, proclaim, educated \\ \hline
    {\em story} & heartbreaking, frightening, terrifying, horrifying, awful, painful, triggering, insightful \\ \hline
    {\em assault} & prevention, policing, devaluing, mishandling, payouts, regrettable \\ \hline
    {\em sex} & perform, oral, consensual, date, violent, nonconsensual \\ \hline
     {\em harassment} & misconduct, rampant, ubiquity, experiencing, assaults, secrecy \\ \hline
    {\em \#metoo} & \#spite, \#mentalhealth, \#gossip, \#sexpredator, \#activism, investigative\\ \hline
    {\em movement} & travesty, witchhunt, hysterical, concerns, nonsense, ridiculously, damaging \\ \hline
    {\em timesup} & oprahs, deathknell, globes, gowns, attendees, \#golden, staged \\ \hline
    {\em woman} & single, unconscious, humiliated, dragged, publicly, qualified, intimidated, backed \\ \hline
    {\em abuse} & exploitation, stigma, secrecy, psychological, admitting, harassment, severity \\ \hline
    {\em victims} & survivors, condemning, offenders, assistance, minimize, pedophiles, bystanders, prevent, suffering \\ \hline
    {\em rape} & attempted, kits, marital, molestation, hookup, aggression, shame \\ \hline
    \end{tabular}
    \caption{Semantic and syntactic co-occurrence patterns from tweets. Keywords are the 12 most frequent words. The right column shows the most co-occurred words associated with left.}
    \label{tab:wordvec}
    \vspace{-28pt}
    \end{table}

	\vspace{-20pt}
	\section{Individual Emotions through Linguistic Markers}
    \vspace{-10pt}
	\subsection{Emotion Attributes}
	\vspace{-7pt}
	We use the psycholinguistic lexicon LIWC\footnote{http://liwc.wpengine.com/} to characterize and compare the type of emotions expressed on both platforms. We obtain measures of the attributes related to user behavior: emotionality (how people are reacting including insight, sad, anger, anxiety, positive and negative emotions), social relationships (family), and individual differences (work, bio, death, swear, sexual). For each attribute, we use $t$-test to check if the Twitter distribution is not significantly different from those of Reddit and the null hypothesis is rejected if p-value$\leq 0.05$.
  
\begin{wrapfigure}{r}{0.5\textwidth}
    \vspace{-10mm}
		\begin{center}
		\includegraphics[width=0.5\textwidth]{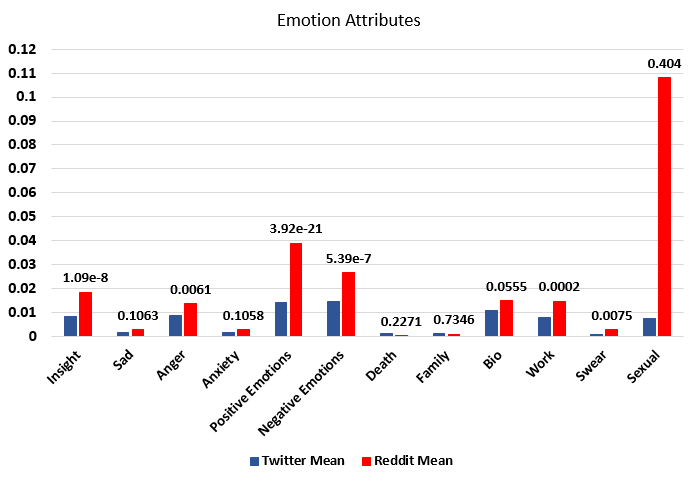}
        \end{center}
        \vspace{-7mm}
		\caption{\textbf{Emotion attributes for Reddit and Twitter posts. Numbers on the bar show the p-value.}}\label{tab:LIWC}
        \vspace{-25pt}
     \end{wrapfigure}

	Figure~\ref{tab:LIWC} shows that the distribution of insight, anger, work, swear and positive and negative emotions attributes in Twitter is significantly different from those of Reddit. In contrast, posts on both platforms have the same distribution of sadness, anxiety attributes. Moreover, the distribution of death, family, bio and sexual-related posts are not significantly different in both platforms.
	\vspace{-18pt}
	\subsection{Sentiment Extraction}
\begin{figure}[ht]\vspace{-25pt}
		\centering 
		\subfloat[\bf{Reddit}]{\includegraphics[width=0.4\textwidth]{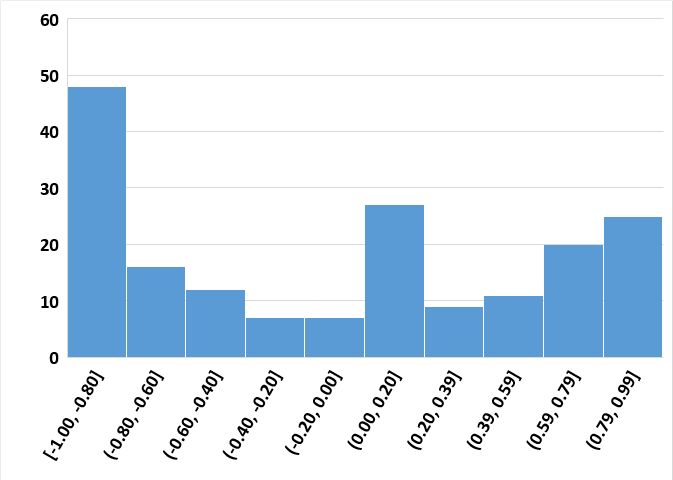}}\quad \quad \quad \quad
		\subfloat[\bf{Twitter}]{\includegraphics[width=0.4\textwidth]{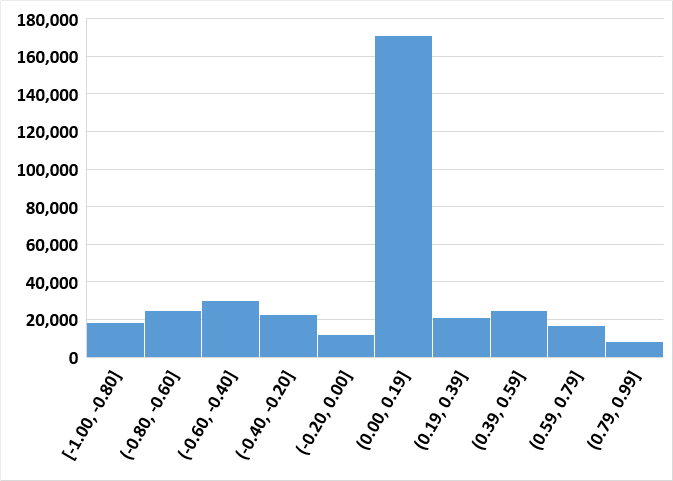}}
        \vspace{-2mm}
		\caption{\textbf{Sentiment distribution for Reddit and Twitter posts}}\label{fig:Sentiment}
        \vspace{-25pt}
	\end{figure}

    \vspace{-5pt}
	To measure the type of sentiment on both platforms, we use Vader~\cite{gilbert2014vader} -- a sentiment analysis tool designed specifically to extract sentiments from social media posts. Results are shown in Figure~\ref{fig:Sentiment} with the following observations. Reddit posts are generally more negative than Twitter posts. This might be because people have no limitation on their posts lengths and thus can easily share their feelings about their stories. However, few posts on Reddit express positive sentiment emphasizing to support the movement. Considering these platforms exclusively, the ratio of positive to negative posts on these platforms are equal to each other showcasing the presence of positivity towards the movement. 
    \vspace{-13pt}
	\section{Conclusions}
    \vspace{-10pt}
    Social media is enabling individuals to recognize the importance of addressing the highly stigmatized issues such as sexual abuse. The research presented in this paper focuses on this particular topic by emphasizing on how users on these platforms (Twitter and Reddit specifically) share their own experiences and respond to the experiences shared by other users. Our investigation suggests that Reddit enables individuals to share their personal stories in depth while on Twitter, users tend to pursue other users to continue the \#metoo movement.But it is because of the posts on Twitter, the  \#metoo movement became viral. Irrespective of the negativity towards different aspects of these personal experiences, individuals on these platforms are positively hoping that these stories will bring a real change in the current society. We hope that our work is useful to initiate discussions between the individuals in the society as well as researchers and lawmakers to propose new laws and regulations to protect individuals in the society. 

\end{document}